\documentclass[5p,sort&compress,times,fleqn]{elsarticle}
\usepackage{graphicx}
\usepackage{amssymb}
\usepackage{color}
\usepackage{enumerate}
\usepackage[breaklinks]{hyperref}
\hypersetup{colorlinks,urlcolor=blue,citecolor=red,linkcolor=blue}
\usepackage[latin1,utf8]{inputenc}
\usepackage[OT2,OT1]{fontenc}

\usepackage{latexsym}
\usepackage{amssymb,amsthm}
\usepackage{bm}

\newcommand{\beaa}{\begin{equation} \begin{array}{ll}}
\newcommand{\eeaa}{\end{equation} 	\end{array} }
\newcommand{\Eq}[1]{(\ref{#1})}

\def\cob{\color{blue}}

\newcommand{\book}[5]{{#1}, #2, #3, #4, #5}
\newcommand{\books}[4]{{#1}, #2, #3, #4}
\newcommand{\oarX}[1]{\href{http://arxiv.org/abs/#1}{{\ttfamily\cob arXiv:#1}}}
\newcommand{\arX}[1]{\href{http://arxiv.org/abs/#1}{{\ttfamily\cob arXiv:#1}}}
\newcommand{\doin}[6]{\href{http://dx.doi.org/#1}{{\cob #2 #3 {#4} (#6) #5}}}
\newcommand{\doinn}[5]{\href{http://dx.doi.org/#1}{{\cob #2 {#3} (#5) #4}}}
\newcommand{\doij}[5]{\href{http://dx.doi.org/#1}{{\cob #2 #3 (#5) #4}}}

\newcommand{\procsinm}[5]{in \emph{#1}, Ed.\ by #2, #3, #4, #5}
\newcommand{\tia}[1]{{#1},}

\newcommand{\be}{\begin{eqnarray}}
\newcommand{\ee}{\end{eqnarray}}
\newcommand{\ba}{\begin{eqnarray}}
\newcommand{\ea}{\end{eqnarray}}

\usepackage{tikz}

\def\b{\beta}
\def\de{\delta}
\def\g{\gamma}

\def\k{\kappa}

\def\ve{\varepsilon}

\def\G{\Gamma}

\def\vr{\varrho}

\def\N{\nabla}

\def\cJ{\mathcal{J}}
\def\cK{\mathcal{K}}

\def\cO{\mathcal{O}}
\def\cP{\mathcal{P}}

\def\ds{d_{\rm S}}
\def\dh{d_{\rm H}}

\def\p{\partial}

\def\cob{\color{blue}}

\def\lp{\ell_{\rm Pl}}

\def\rmd{d}

\begin{document}

\begin{frontmatter}

\title{
Gravitational-wave luminosity distance in quantum gravity
}

\author{Gianluca Calcagni}
\address{Instituto de Estructura de la Materia, CSIC, Serrano 121, 28006 Madrid, Spain}

\author{Sachiko Kuroyanagi}
\address{Department of Physics, Nagoya University, Chikusa, Nagoya 464-8602, Japan}
\address{Instituto de F\'isica Te\'orica UAM-CSIC, Universidad Auton\'oma de Madrid, Cantoblanco, 28049 Madrid, Spain}

\author{Sylvain Marsat}
\address{APC, AstroParticule et Cosmologie, Universit\'{e} Paris Diderot, CNRS/IN2P3, CEA/Irfu, Observatoire de Paris, Sorbonne Paris Cit\'{e}, 10, rue Alice Domon et L\'{e}onie Duquet 75205 PARIS Cedex 13, France} 

\author{Mairi Sakellariadou}
\address{Theoretical Particle Physics and Cosmology Group, Physics Department, King's College London, University of London, Strand, London WC2R 2LS, United Kingdom}

\author{Nicola Tamanini}
\address{Max-Planck-Institut f\"{u}r Gravitationsphysik, Albert-Einstein-Institut, Am M\"{u}hlenberg 1, 14476 Potsdam-Golm, Germany}

\author{Gianmassimo Tasinato}
\address{Department of Physics, Swansea University, Swansea, SA2 8PP, UK}

\begin{abstract}
Dimensional flow, the scale dependence of the dimensionality of spacetime, is a feature shared by many theories of quantum gravity (QG). We present the first study of the consequences of QG dimensional flow for the luminosity distance scaling of gravitational waves in the frequency ranges of LIGO and LISA. We find generic modifications with respect to the standard general-relativistic scaling, largely independent of specific QG proposals. We constrain these effects using two examples of multimessenger standard sirens, the binary neutron-star merger GW170817 and a simulated supermassive black-hole merger event detectable with LISA. We apply these constraints to various QG candidates, finding that the quantum geometries of group field theory, spin foams and loop quantum gravity can give rise to observable signals in the gravitational-wave spin-2 sector. Our results complement and improve GW propagation-speed bounds on modified dispersion relations. Under more model-dependent assumptions, we also show that bounds on quantum geometry can be strengthened by solar-system tests.
\end{abstract}
\end{frontmatter}



\date{March 28, 2019}

\section {Introduction}

Quantum gravity (QG) includes any approach aiming at unifying General
Relativity (GR) and quantum mechanics consistently, so as to keep
gravitational ultraviolet (UV) divergences under control
\cite{Ori09,Fousp}. Any such approach can be either top-down or
bottom-up, depending on whether it prescribes a specific geometric
structure at the Planck scale, or it starts from low energies and then
climbs up to higher energy scales. The former class includes string
theory, nonlocal QG, and  nonperturbative proposals as
Wheeler--DeWitt canonical gravity, loop QG, group field
theory, causal dynamical triangulations, causal sets, and
noncommutative spacetimes. The latter class contains asymptotic safety
and the spectral approach to noncommutative geometry. Such variety of QG theories
 leads to many cosmological consequences which are currently under
investigation \cite{CQC}. 

Given the recent direct observations of gravitational waves (GW)
\cite{Ab16a,Ab16b,Ab16c,Ab17a,Ab17b,Abb18,LIGOScientific:2018mvr},
opening a new era in GW and multimessenger astronomy, 
new opportunities are arising to test theories beyond GR. In general, QG may affect both the
\emph{production} \cite{YYP,BYY} and the \emph{propagation} of GWs
\cite{YYP,EMNan,ArCa2,MYW} in ways that differ from
those obtained from modified-gravity models for dark energy. While
QG aims at regularizing UV divergencies in a framework
applying the laws of quantum mechanics to the gravitational force,
one might hope that
yet-to-be developed connections between UV and infrared
 regimes of gravity can
  lead to a consistent theory of dark energy from QG.

On one hand, one may believe that QG theories can leave no signature in GWs,
arguing that quantum effects  will be suppressed by the Planck
scale. Such a conclusion is reached by considering the leading-order
perturbative quantum corrections to the Einstein--Hilbert
action. Since these corrections are quad\-rat\-ic in the curvature and
proportional to the Planck scale $\lp\approx 10^{-35}\,{\rm m}=5
\times 10^{-58}\,{\rm Mpc}$, they are strongly subdominant at
energy or curvature scales well above $\lp$. For instance, for a
Friedmann--Lema\^itre--Robertson--Walker (FLRW) universe, there are
only two scales for building dimensionless quantities:
$  \lp$ and the Hubble radius $H^{-1}$. Therefore, quantum contributions  
should be of the form $(\lp H)^n$, where $n=2,3,\dots$. Today, quantum
effects  are as small as $(\lp H_0)^n\sim 10^{-60 n}$, and any
late-time QG imprint is Planck-suppressed and undetectable.

On the other hand, these considerations are not necessarily correct. One may consider nonperturbative effects going beyond the simple
dimensional argument quoted above. Indeed, in the presence of a third
intermediate scale $L\gg\lp$, quantum corrections may become
$\sim\lp^a H^b L^c$ with $a-b+c=0$, and not all these exponents are
necessarily small. Such is the case, for instance, of loop
quantum cosmology with anomaly cancellation (a mini-superspace model
motivated by loop quantum gravity), where quantum states of spacetime
geometry may be endowed with a mesoscopic effective scale \cite{BCT1}. These and other QG inflationary models can leave a sizable imprint in the early universe \cite{CQC}. 

In this Letter, we consider a long-range nonperturbative mechanism, {\sl
  dimensional flow}, namely the change of spacetime dimensionality found in
most QG candidates \cite{tH93,fra1,Car17}. We argue that
this feature of QG, already used as a direct agent in QG inflationary models \cite{RSnax,AAGM1,AAGM3,frc14}, can also have important consequences for the
propagation of GWs over cosmological distances.
We identify QG predictions shared by different quantization schemes, and
  determine a model-independent expression, Eq.\ \Eq{dla}, for
the luminosity distance of GWs propagating in a dimensionally
  changing spacetime in QG. Testing this expression against current
LIGO-Virgo data, mock LISA data, and solar-system tests, allows us to
constrain the spacetime dimensionality of a representative number of
 QG theories. We mainly focus on 
  the spin-2 GW sector and on specific opportunities of GW experiments
  to test QG scenarios, assuming that the other dynamical sectors (e.g.~spin-0 and spin-1) are not modified by QG corrections. 
   Our results suggest that group field theory/spin foams/loop quantum gravity (GFT/SF/LQG), known to affect both the UV limit of gravity and cosmological inflationary scales, can also influence the  
   properties of GWs, due to effects that have not been previously considered.  
    We also compare our results with complementary constraints on modified dispersion relations, 
    and discuss possible implications of the Hulse--Taylor pulsar. 
Finally, we also take into consideration some different type of model-dependent bounds to QG theories, particularly from solar-system experiments.

\section{Dimensional flow}


All the main QG theories share some features that will be the basis for our results. In general, there always exists a continuum limit to a spacetime with a continuous integrodifferential structure, effectively emerging from some fundamental dynamics that we do not need to specify here. On this continuum, one can consider a gravitational wave, which, in Isaacson shortwave approximation \cite{Isaacson:1967zz}, is a high-frequency spin-2 perturbation $h_{\mu\nu}=h_+e^+_{\mu\nu}+h_\times e^\times_{\mu\nu}$ over a background metric $g^{(0)}_{\mu\nu}=g_{\mu\nu}-h_{\mu\nu}$ and is described by the two polarization modes $h_{+,\times}$ in $D=4$ topological dimensions (with $e^{+,\,\times}_{\mu\nu}$ being the polarization tensors). Quantization of spacetime geometry or its emergence from fundamental physics introduces, directly or indirectly, two types of change relevant for the propagation of GWs: an anomalous spacetime measure $\rmd\vr(x)$ (how volumes scales) and a kinetic operator $\cK(\p)$ (modified dispersion relations). Other effects such as perturbative curvature corrections are not important here. The perturbed action for a small perturbation $h_{\mu\nu}$ over a background $g^{(0)}_{\mu\nu}$ is
\be\label{hboxh2} \hspace{-.5cm}
S=\frac{1}{2\ell_*^{2\Gamma}}\!\!\int\!\rmd\vr\sqrt{-g^{(0)}}
\left[h_{\mu\nu}\cK h^{\mu\nu}\!+O(h_{\mu\nu}^2)+\cJ^{\mu\nu}h_{\mu\nu}\right]\!, 
\ee
 where the prefactor makes the action dimensionless, $\cJ^{\mu\nu}$ is a generic source term, 
 and the $O(h_{\mu\nu}^2)$ terms play no role at small scales.
 The modes $h_{+,\times}/\ell_*^{\Gamma}$, where $\ell_*$ is a
 characteristic scale of the geometry, are dimensionally and
 dynamically equivalent to a scalar field.
	
The measure defines a geometric observable, the Hausdorff dimension $\dh(\ell):={\rmd\ln\vr(\ell)}/{\rmd\ln\ell}$,
describing how volumes scale with their linear size $\ell$.
In a classical spacetime, $\dh=4$. 
Also, spacetime is dual to a well-defined momentum space
characterized by a measure $\tilde\vr(k)$ with Hausdorff
dimension $\dh^k$, in general different from $\dh$. The kinetic term
is related to $\dh^k$ and to another observable, the
spectral dimension $\ds(\ell):=-{\rmd\ln\cP(\ell)}/{\rmd\ln\ell}$,
where 
$\cP(\ell)\propto\int\tilde\vr(k)\,\exp[-\ell^2\tilde\cK(-k^2)]$, 
and the function $\tilde\cK$ 
is the dispersion relation $\cK$ rescaled by a length power. In any plateau of dimensional flow, where all dimensions are approximately constant, hence $\tilde\vr(k)\sim\rmd k\,k^{\dh^{k}-1}$ and $\tilde\cK\simeq \ell_*^{2\b-2}k^{2\b}$ for a constant $\b=[\cK]/2$ (half the energy scaling of $\cK$), we find that $\cP\propto(\ell_*^{\b-1}\ell)^{-\dh^{k}/\b}$, implying $\ds=2\dh^k/[\cK]$. In such plateau
region, since $[S]=0$, from Eq.\ \Eq{hboxh2} we have  
\be\label{vp}
\Gamma\simeq\frac{\dh}{2}-\frac{\dh^k}{\ds}\,,
\ee
and $\Gamma\approx{\rm const}$.
We assume that $\ds\neq 0$ at all scales. Cases where $\ds=0$ at short scales must be treated separately \cite{BrCM}. In the GR limit in $D$ topological dimensions (standard spacetime, no QG corrections), $\dh=\dh^k=\ds=D$ and $\Gamma=D/2-1$, the usual scaling of a scalar field. 
 
 Equation \Eq{vp} applies to many concrete QGs, each with its own characteristic measures $\vr$, $\tilde\vr$ and kinetic operator $\cK$. Predictions of representative theories at small ($\Gamma_{\rm UV}$) and intermediate scales ($\Gamma_{\rm meso}$) are found in Tab.\ \ref{tab1}. Scales at which QG corrections are important belong to the UV regime, whereas intermediate scales where the corrections are small but non-negligible belong to the mesoscopic one.
\begin{table*}[ht!]
\begin{center}
\begin{tabular}{|c|c|c|}\hline		 
																						 &$\Gamma_{\rm UV}$ & $\Gamma_{\rm meso}\gtrsim 1$\\\hline\hline

GFT/SF/LQG \cite{ACAP,COT3,MiTr} 	 &$[-3,0)$ & {\rm yes} \\

Causal dynamical triangulations  \cite{CoJu}	       &$-2/3$ & \\

$\k$-Minkowski (other) \cite{Ben08,ArTr} 
 &  $[-1/2,1]$ & \\
Stelle gravity \cite{Ste77,CMNa}    														   & 0      & \\
String theory (low-energy limit) \cite{ACEMN,CaMo1}					 &0      & \\
Asymptotic safety \cite{LaR5}    														 &  0      & \\
Ho\v{r}ava--Lifshitz gravity \cite{Hor3}										 &  0      & \\
$\k$-Minkowski bicross-product $\N^2$ \cite{ArTr}
   & $3/2$ & yes \\
$\k$-Minkowski relative-locality $\N^2$ \cite{ArTr}
   &  2		  & yes \\
Padmanabhan nonlocal model \cite{Pad98,ArCa1}	      				   & 2 & yes \\
\hline
\end{tabular}
\end{center}
\caption{\label{tab1}  {\it Value of $\Gamma_{\rm UV}$ for different QG theories. Theories with a near-IR parameter $\Gamma_{\rm meso}\gtrsim 1$ are indicated in the second column.}}
\end{table*}

Given a spacetime measure $\vr$, a kinetic operator $\cK$, and a compact source $\cJ$, the GW amplitude $h$ (subscripts $+,\times$ omitted) is determined by the convolution $h\propto\int\rmd\vr\,\cJ\,G$ of the source with the retarded Green function obeying $\cK\,G=\de_\vr$, where $\de_\vr$ is the Dirac delta generalized to a nontrivial measure $\vr$. In radial coordinates in the local wave zone (a region of space larger than the system size, but smaller than any cosmological scale), $G(t,r)\sim f_G(t,r)\,r^{-\G}$, where $f_G$ is dimensionless. This yields the scaling of $h$,
\be\label{hr} 
h(t,r)\sim f_h(t,r)\,\left({\ell_*}/{r}\right)^{\G},\qquad [f_h]=0\,. 
\ee
Equation \Eq{hr} describes the distance scaling of the amplitude of GW radiation emitted by a binary system and observed in the local wave zone, in any regime where $\G\approx{\rm const}$. $f_h$ depends on the source $\cJ$ and on the type of correlation function (advanced or retarded), but the key point is that $h$ is the product of a dimensionless function $f_h$ and a power-law distance behavior. This is a fairly general feature  in QG, since it is based only on the scaling properties of the measure and the kinetic term.


\section{Gravitational waves}

We now extend these results to GWs
propagating over cosmological distances. Working on a conformally flat FLRW
 background, $t\to\tau$ is conformal time, $r$ is the comoving distance of the
 GW source from the observer, and $r$ is multiplied by the scale
 factor $a_0=a(\tau_0)$ in the right-hand side of Eq.\ \Eq{hr}. To express Eq.~\Eq{hr} in terms of an observable, we
 consider GW sources with an electromagnetic counterpart.
 The luminosity distance of an object emitting
 electromagnetic radiation is defined as the power L per flux unit F,
 $d_L^\textsc{em}:=\sqrt{{\rm L}/(4\pi {\rm F})}$, and it is measured photometrically. On a flat FLRW
 background, $d_L^\textsc{em}=(1+z)\int^{\tau_0}_{\tau(z)}\rmd
 \tau=a_0^2r/a$,
 where $z=a_0/a-1$ is the
 redshift. We assume that QG corrections to $d_L^\textsc{em}$ 
 are negligible at
 large scales. Absorbing redshift factors and all the details of the
   source (chirp mass, spin, and so on) into the dimensionless function $f_h(z)$, Eq.\ \Eq{hr} becomes
\be\label{h1} 
h(z)\sim f_h(z)\,\left[\frac{\ell_*}{d_L^\textsc{em}(z)}\right]^{\G}. 
\ee 

The final step is to generalize relation (\ref{h1}), valid only for a plateau in dimensional flow, to all scales. An exact calculation is extremely difficult except in special cases, but a model-independent approximate generalization is possible because the system is \emph{multiscale} (it has at least an IR and a UV limit, $\G\to 1$ and $\G\to\G_{\rm UV}$). In fact, multiscale systems such as those in multifractal geometry, chaos theory, transport theory, financial mathematics, biology and machine learning are characterized by at least two critical exponents $\G_1$ and $\G_2$ combined together as a sum of two terms $r^{\G_1}+ A \, r^{\G_2}+\dots$, where $A$ and each subsequent coefficients contain a scale (hence the term multiscale). In QG, lengths have exactly this behavior, which
 has been proven to be 
  universal \cite{NgDa,Ame94,first,ACCR,CaRo} in the flat-space limit: it must hold also for the luminosity distance because one should recover such multiscaling feature  in the subcosmological limit $d_L^\textsc{em}\to r$. Thus,
\be\label{dla} 
h \propto \frac{1}{d_L^\textsc{gw}}\,,\qquad \frac{d_L^\textsc{gw}}{d_L^\textsc{em}}=1+\ve
\left(\frac{d_L^\textsc{em}}{\ell_*}\right)^{\g-1}, 
\ee 
with $\ve={\cal O}(1)$, and $\gamma\neq 0$. In the presence of only one fundamental length scale
$\ell_*=\cO(\lp)$, Eq.\ \Eq{dla} is exact \cite{first} and $\g=\Gamma_{\rm UV}$ takes the values in Tab.\ \ref{tab1}. Conversely, if $\ell_*$ is a mesoscopic scale, then Eq.\ \Eq{dla} is valid only near the IR, close to the
  end of the flow, and $\g=\Gamma_{\rm meso}\approx 1$.
  
 The coefficient $\ve$ cannot be determined universally, since it depends on the details of the
  transient regime, but we can set $\ve=\cO(1)$ without loss of generality because also $\ell_*$ is a free parameter. However,
  the case with $\g\approx 1$ is subtle as we cannot recover
  GR unless $\ve$ vanishes. This implies that $\ve$ must
  have a $\g$ dependence: the simplest   choice such that $\ve(\g\neq 1)=\cO(1)$, $\ve(\g=1)=0$, and
  recovering the pure power law \Eq{h1} on any plateau is $\ve=\g-1$. The sign of $\ve$ is left undetermined to allow for all possible cases. The
     result is Eq.~\Eq{dla} with $\ve=\pm|\g-1|$.

   Equation \Eq{dla} is our key result for analyzing the phenomenological consequences of QG dimensional
   flow for the propagation of GWs. Its structure resembles the GW luminosity-distance relation expected in some models with large
    extra-dimensions
    \cite{DeMe,PFHS,Andriot:2017oaz,Abb18}, where gravity
    classically ``leaks'' into a higher dimensional space. 
 However, we emphasize that Eq.\ \Eq{dla} is based on a feature of
    most QG proposals, dimensional flow, and does not rely on
    realizations in terms of classical extra dimensions.

The left-hand side of Eq.\ \Eq{dla} is the strain measured in a GW
interferometer. The right-hand side features the luminosity distance
measured for the optical counterpart of the standard siren. Therefore,
observations can place constraints on the two parameters $\ell_*$ and
$\g$ in a model-independent way, by constraining the ratio
$d_L^\textsc{gw}(z)/d_L^\textsc{em}(z)$ as a function of the redshift
of the source. Our  analysis is based on two standard sirens (with associated EM counterpart): the
binary neutron-star merger GW170817 observed by LIGO-Virgo and the Fermi
telescope \cite{Ab17b}, and a simulated $z=2$ supermassive black
hole merging event that could be observed by LISA \cite{CKMSTT,Tamanini:2016zlh,Tamanini:2016uin}. 
There are three cases to consider:

(a) $0>\g-1$ leads to an upper bound on $\ell_*$ of
  cosmological size, namely $\ell_*<(10^1-10^4)\,{\rm Mpc}$. Hence, when $\g=\G_{\rm UV}$, \emph{we cannot
   constrain the deep UV limit of quantum gravity}, since $\ell_*={\cal O}(\lp)$. This is
  expected in QG theories with $\G_{\rm UV}<1$
  (Tab.~\ref{tab1})
   on the  tenet that
  deviations from classical geometry occur at microscopic scales
  unobservable in astrophysics.

(b) $0<\g-1={\cal O}(1)$: there is a lower bound on $\ell_*$ of
  cosmological size. Therefore, if Eq.\ \Eq{dla} is interpreted as
  valid at \emph{all} scales of dimensional flow and $\g=\G_{\rm UV}$, this
  result rules out the three models not included in the
  previous case: $\k$-Minkowski spacetime with ordinary measure and
  the bicross-product or relative-locality Laplacians and
  Padmanabhan's nonlocal model of black holes.

\begin{figure}[ht!]
\centering
\includegraphics[width=8.5cm]{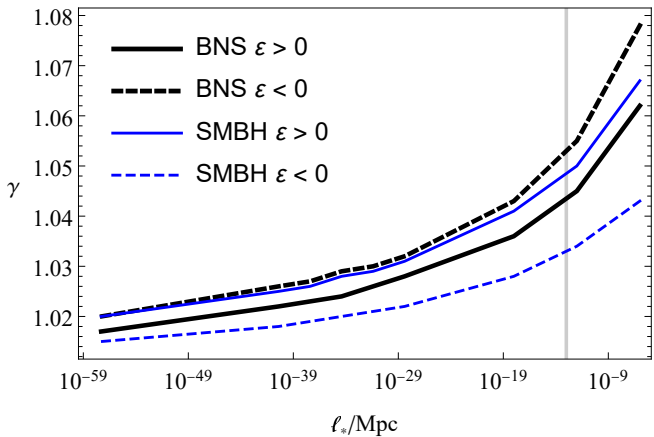}
\caption{\label{fig1} \it Upper bounds on $\g$ for $\ell_*$ fixed
  between $1\,{\rm Mpc}$ and the Planck scale $\lp=5\times
  10^{-58}\,{\rm Mpc}$ for the LIGO-Virgo observed binary neutron-star merger
  GW170817 (BNS) and a simulated LISA supermassive black hole (SMBH) merger.}
\end{figure}

(c) $0<\g-1\ll 1$: Eq.\ \Eq{dla} is valid in a near-IR regime and
  $\g=\G_{\rm meso}$ is very close to 1 from above. Using a Bayesian analysis identical to that of \cite{Abb18} (page 11) where $\ell_{*}$ is fixed and the constraint on $\g$ is inferred \cite{CKMSTT}, the resulting
  upper bound on $\g$ is shown in Fig.~\ref{fig1}. 
For the smallest QG scales, the bound saturates to
\be\label{bou1}
0\,<\,\Gamma_{\rm meso}-1\,<\,0.02~.
\ee
Examining Eq.~\Eq{vp}, we conclude that case (c)
is realized only for geometries with a spectral dimension reaching $\ds\to
4$ from above. The only theories in our list that do so are those
where $\G_{\rm UV}>\G_{\rm meso}>1$ (the last three in Tab.\ \ref{tab1}: $\k$-Minkowski spacetime with
ordinary measure and bicross-product or relative-local\-i\-ty Laplacians
and Padmanabhan's model \cite{CKMSTT}) 
or $\G_{\rm meso}>1>\G_{\rm UV}$ (GFT/SF/LQG \cite{COT3}).
However, we exclude observability of the 
models with $\G_{\rm UV}>\G_{\rm meso}>1$, since they predict $\G_{\rm meso}-1\sim (\lp/d_L^\textsc{em})^2< 10^{-116}$
\cite{CKMSTT}.
Thus, only GFT, SF or LQG could generate a signal
detectable with standard sirens.
Here $\ds$
runs from small values in the UV, but before reaching
the limit $\ds^{\rm IR}=4$ it overshoots the asymptote and decreases again: hence
$\G_{\rm meso}>1>\G_{\rm UV}$. It would be interesting to find
realistic quantum states of geometry giving rise to such a signal, with the
 construction of simplicial complexes as in Ref.~\cite{COT3}.


\section{Complementary constraints}

Dimensional flow is also influenced by modifications of the dispersion relation $\cK(-k^2)=-\ell_*^{2-2\dh^k/\ds}k^2+k^{2\dh^k/\ds}$ of the spin-2 graviton field, and this fact has been used to impose constraints on QG theories exhibiting dimensional flow 
using the LIGO-Virgo merging events \cite{YYP,EMNan,ArCa2}. However, the limits obtained this way are weaker than the ones we have found here because the GW frequency is much lower than the Planck frequency.
One gets either very weak bounds on $\ell_*$ or, setting $\ell_*^{-1}>10\,{\rm TeV}$ (LHC scale), a bound $n=\dh-2-2\G<0.76$ \cite{ArCa2}, for $\dh^{\rm meso}\approx 4$ corresponding to $\G_{\rm meso}-1>-0.38$. This can constrain models such as the second and third in Tab.~\ref{tab1}, but not those such as GFT/SF/LQG for which Eq.~\Eq{bou1} holds. 

Additional constraints on the spin-2 sector can arise from observations of the Hulse--Taylor pulsar \cite{WeTa}.
If the spacetime dimension deviates from four roughly below scales $l_{\rm pulsar}=10^6\,{\rm km}\approx 10^{-13}\,{\rm Mpc}$, then the GW emission from this source is expected to be distinguishable from GR.
However, it is  difficult to analyze the binary dynamics and GW emission in 
higher-dimensional spacetimes \cite{CDL} and it is
consequently more complicated to set bounds from binary pulsar systems.
We will thus leave these investigations for future work.
We point out, however, that at scales below $\ell_*=l_{\rm pulsar}$ (the vertical line in Fig.~\ref{fig1}),
our results could be largely improved by stronger constraints from the dynamics of compact objects.

Finally, stronger but model-dependent bounds can arise in scenarios that
affect other sectors besides the dynamics of the spin-2 graviton field.
%
      To have an idea of the constraints that can arise when other sectors become dynamical in QG, we consider
    a case where
  the 
   effective scalar Newtonian
  potential $\Phi\sim h_{00}$ experiences QG dimensional flow: then
  the bound \Eq{bou1} can be strengthened by solar-system
  tests.
In fact, Eq.\ \Eq{hr} can describe $\Phi$
in a regime where $\G$ is approximately constant, while choosing
subhorizon distances $d_L^{\textsc{em}}=r$ in Eq.\ \Eq{dla} we
get a multiscale expression. Thus, in four  
dimensions 
\be\label{newt}
 \hspace{-.5cm}\Phi\propto -
\frac1r\left(1\pm\frac{\Delta\Phi}{\Phi}\right),\qquad
\frac{\Delta\Phi}{\Phi}=|\gamma-1|\left(\frac{r}{\ell_*}\right)^{\gamma-1}.
\ee 
This result, different from but complementary \cite{CKMSTT}
  to what found in the effective field theory approach to QG, applies to the
  nonperturbative GFT/SF/LQG theories with $\g>1$ at mesoscopic
  scales. Assuming that photon geodesics are not modified at those
scales, GR tests within the solar system using the Cassini bound
impose $\Delta\Phi/\Phi< 10^{-5}$ \cite{BIT,Wil14}, implying 
\be
0<\Gamma_{\rm meso}-1<10^{-5}, 
\ee 
which is stronger than the limit obtained from GWs. However, this result relies on model-dependent assumptions on the scalar sector, independent of our previous arguments on the propagation of spin-2 GWs, and should be taken \emph{cum grano salis}. We emphasize that in QG the dynamics of spin-0 fields and the Newtonian potential $\Phi$ can be far from trivial. Precisely for GFT/SF/LQG, the classical limit of the graviton propagator is known \cite{BMRS}, but corrections to it and to the Newtonian potential are not \cite{CLS}. Therefore, we cannot compare Eq.\ \Eq{newt} with the full theory, nor do we know whether quantum states exist giving rise to such a correction.

\section{Conclusions}


Quantum gravity can modify both the production and the propagation of
gravitational waves. We obtained the general equation \Eq{dla} describing model-independent modifications due to nonperturbative QG
on the GW luminosity distance associated with long distance propagation of GWs. We have then shown that, while the deep UV regime of QG cannot be probed by GWs, mesoscopic-scale (near-IR) departures from classical GR due to QG effects can be in principle testable with LIGO and LISA detections of merging events in the theories GFT/\-SF/\-LQG.
Solar-system tests of the Newtonian potential $\Phi$ lead to
stronger constraints than the ones imposed from GW
data, but rely on model-dependent assumptions on
the dynamics of the scalar Newtonian potential  $\Phi$. Focussing  on the spin-2 field only, there are several directions that remain to be
explored. For instance, time delays in gravitational lensing might be
another place where to look for propagation effects beyond GR within
LISA sensitivity. Moreover, also the details of the astrophysical
systems giving rise to GW signals should be studied, in order to
understand the consequences of a QG geometry on the
production of GWs in the high-curvature region surrounding compact
objects.

\medskip

\noindent
\emph{Acknowledgments.} 
G.C.\ and S.K.\ are supported by the I+D grant FIS2017-86497-C2-2-P of the Spanish Ministry of Science, Innovation and Universities. S.K.\ is supported by JSPS KAKENHI No.~17K14282 and Career Development Project for Researchers of Allied Universities. M.S.\ is supported in part by 
 STFC grant 
ST/P000258/1. G.T.\ thanks Ivonne Zavala for discussions; he is partially supported by STFC grant ST/P00055X/1.


\end{document}